\documentclass[12pt,a4paper]{article}
  \usepackage[margin=5mm,a5paper]{geometry}

\title{Surface deformation and shear flow in ligand mediated cell adhesion}

\author{
Sarthok Sircar\thanks{School of Mathematical Sciences, University of Adelaide, South Australia~5005, Australia. Corresponding author \protect\url{mailto:sarthok.sircar@adelaide.edu.au}} 
\and
Anthony J. Roberts\thanks{School of Mathematical Sciences, University of Adelaide, South Australia~5005, Australia. 
\protect\url{mailto:anhtony.roberts@adelaide.edu.au}}
}

\usepackage{microtype, amsmath, amssymb, url, subfigure}
\usepackage{graphicx,bm}
\usepackage[left]{lineno}

\renewcommand{\vec}[1]{\text{\boldmath$#1$}}
\newcommand{\ben}{\begin{equation}}
\newcommand{\een}{\end{equation}}

\begin{document}
\maketitle

\begin{abstract}
We present a single, unified, multi-scale model to study the attachment\slash detachment dynamics of two deforming, near spherical cells, coated with binding ligands and subject to a slow, homogeneous shear flow in a viscous fluid medium. 
The binding ligands on the surface of the cells experience attractive and repulsive forces in an ionic medium and exhibit finite resistance to rotation via bond tilting. 
The macroscale drag forces and couples describing the fluid flow inside the small separation gap between the cells, are calculated using a combination of methods in lubrication theory and previously published numerical results. 
For a select range of material and fluid parameters, a hysteretic transition of the sticking probability curves between the adhesion and fragmentation domain is attributed to a nonlinear relation between the total microscale binding forces and the separation gap between the cells. 
We show that adhesion is favored in highly ionic fluids, increased deformability of the cells, elastic binders and a higher fluid shear rate (until a critical value). 
Continuation of the limit points predict a bistable region, indicating an abrupt switching between the adhesion and fragmentation regimes at critical shear rates, and suggesting that adhesion of two deformable surfaces in shearing fluids may play a significant dynamical role in some cell adhesion applications.

\end{abstract}


\section{Introduction} \label{sec:intro}
The adhesion and fragmentation of cells in suspension is an ubiquitous and biologically significant process.
Examples include binding of bacterial clusters to medical implants or host cell surfaces during infection~\cite{Zhu2000}, cancer cell metastasis~\cite{Basu2013}, coalescence of medical gels with functionalized particles or micro-bubbles for targeted drug delivery~\cite{Sircar2013} and the adherence of platelets and monocytes to atherosclerotic plaques~\cite{Cohen2014}.
Cell adhesion is commonly mediated by specific ligand interactions, e.g., the ligand-mediated surface adhesion is an important case in the experimental studies of the \textsc{p}-selectin/\textsc{psgl}-1 catch bond interactions of leukocytes (a roughly spherical particle) with and without fluid flow~\cite{Marshall2003}.
The adhesive properties of biological surfaces connected by multiple independent tethers are also presently inspiring the development of novel adhesives mimicking the remarkable properties of
beetle and gecko feet~\cite{Varenberg2007}.
Several other applications as well as {in vivo} and {in silico} studies of cell adhesion are listed in Lauffenburger {et al.} \cite{Lauffenburger1993}, Springer \cite{Springer1995}, Hammer {et al.} \cite{Hammer1996}, Jones {et al.} \cite{Jones1996} and Zhu \cite{Zhu2000}. 
However, the models and the experiments listed in these references fail to describe with a single unified theory the several interrelated physical features associated with the adhesion process.
This article develops a unified theory and approach.

The theoretical modelling of the surface adhesion in a fluid-borne environment presents significant challenges.
The adhesive forces are composed of numerous physical processes including ligand-receptor binding kinetics~\cite{Dembo1988}, surface deformation and the related mechanical stresses due to the elastic forces on the cell membrane~\cite{Hodges2002}, excluded volume effects~\cite{Wang2005}, paramagnetism~\cite{Forest2006}, short range interactions~\cite{Sircar2013}, and flow past the surrounding surfaces~\cite{Sircar2009, Sircar2010}, all of which determines the fate of the binding surfaces.
Consequently, many detailed kinetic models have successfully described the adhesion-fragmentation processes from the microscopic perspective.
Schwarz {et al.}~\cite{Korn2006} and more recently Mahadevan {et al.}~\cite{Mani2012} studied the cellular adhesion between the ligand coated wall and a rigid sphere moving in a shear flow.
A similar model by Seifert {et al.}~\cite{Bihr2012} described the membrane adhesion via Langevin simulations.
On the contrary, the macro-scale phase-field models describe the geometry of aggregates as a continuum mass of extracellular polymeric substance and predict the stability of the anisotropic structures in a flowing medium~\cite{Keener2011, Keener2011a}. 
However, a link between the micro-scale and the macro-scale description detailing the several interrelated phenomena in the adhesion process is still missing \cite{MORI2013}, but is now proposed here.

Multi-scale models provide a powerful route to explore possible connections between macroscopic physiological observations, such as the minimum shear threshold for surface adhesion~\cite{Hammer1996} and the microscale mechanochemical effects operating within individual intermolecular bonds~\cite{Dembo1988}.
Sciortino {et al.}~\cite{Corezzi2012} reported research in this direction, but their numerical studies were done with chemically inert particles.
Other examples of recent work includes developing probabilistic extensions of the Smoluchowski's multiplicative aggregation kernel in one~\cite{Odriozola2007} and two dimensions~\cite{Moncho-Jorda2001a}, with kernels containing containing one scaling parameter to be fit to data.
Jia {et al.}~\cite{Jia2006} developed a method for predicting critical coagulant concentration via deriving a kernel incorporating surface charge density and potential as a function of the electrolyte.
Gilbert {et al.}~\cite{Gilbert2007} investigated and validated the forces and potentials for nanoparticles, whereas Babler and Morbidelli~\cite{Babler2007} studied aggregation and fragmentation, but only driven by diffusion and shear flow.
In summary, each of these research efforts have focused on the adhesion and fragmentation using separate theories that we unify.

The aim of this article is to develop and investigate a single, unified, multi-scale (i.e., at the micro-nano level) model for ligand mediated deformable cell-cell adhesion dynamics in a slow, viscous, shear flow conditions. 
This unique study considers several competing physical processes influencing simultaneous transition between the adhesion-fragmentation regimes, namely, binding-unbinding of the ligands, surface deformation, fluid flow and interaction between the charged surface and the liquid medium. 
In the next section, we present the details of the new comprehensive model, including the bond mechanics (\S \ref{subsec:BK}), the interaction of charged surface in a fluid medium (\S \ref{subsec:LRI}), micro-scale binding forces on the cell surface (\S \ref{subsec:MicF}), macro-scale drag forces and couple arising due to the flow-hydrodynamics inside and outside the narrow gap between the cell surface (\S \ref{subsec:MacF}) and the calculation of the adhesion area of deformed cells in slow, viscous shear flow (\S \ref{subsec:AA}). 
The microscale forces and the entire system of non-dimensionalized equations are listed in \S \ref{subsec:ND} and coupled with the macroscale hydrodynamics in \S \ref{subsec:MicMac}. 
Section~\ref{sec:results} highlights the simulation results of the binder kinetics at steady state and the bifurcation analysis in a select range of material and fluid parameters, and conclude (\S \ref{sec:end}) with a brief discussion of the implication of these new results and the focus of our future directions.


\section{Mathematical model: binder kinetics, surface deformation and hydrodynamics} \label{sec:model}

\begin{figure}
\centering
\includegraphics[scale=0.45]{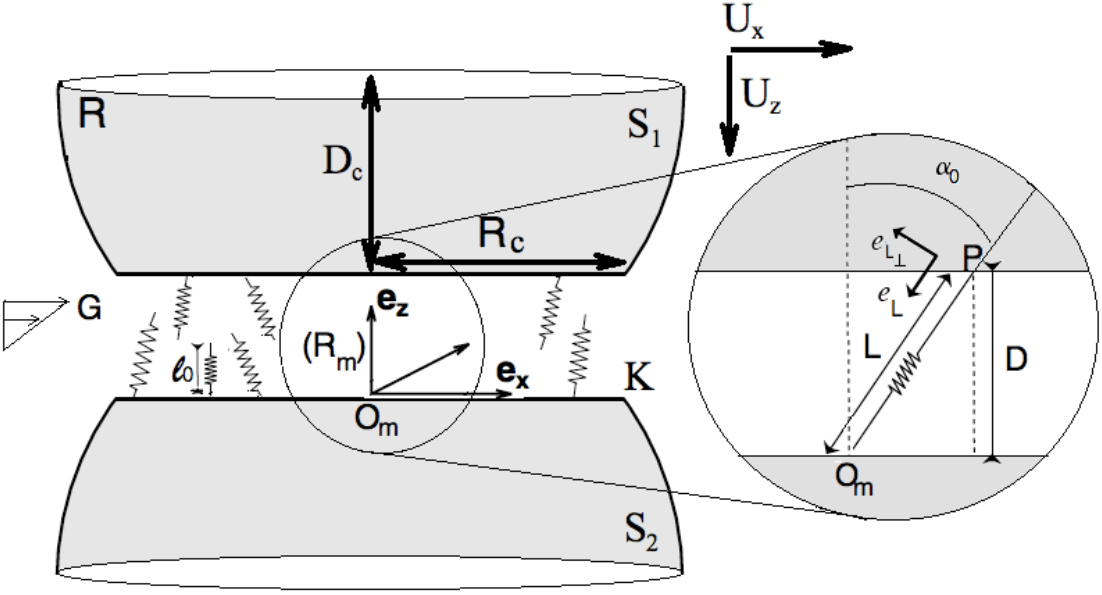}
\caption{An illustration of two spherical, deformable cells coated with binding ligands and translating and rotating in uniform flow.
The symbol~$(R_m)$ denotes the moving frame of reference, with the origin $O_m$ fixed on the surface of sphere~S$_2$ at the centre of the adhesion region.}\label{fig:sphere}
\end{figure}

This section derives the evolution equation governing the dynamics of the binding ligands, attached on the charged surface of the cells and immersed in an electrolytic solvent subject to slow shear flow.
We model a cell as a thin sphere with extensible membrane subject to tension but having negligible bending stiffness, surrounding a fluid interior of fixed volume per unit length.
The viscosity of the cell's interior is assumed to be low enough that it behaves as if it were inviscid.
The effects of gravity, non-specific forces acting on the cell, as well as the roughness of the cell surface are neglected.
Further, we  assume that, subject to a finite tilting, the ligands are fixed on the cell surface.
The spherical cells adhere through well-defined disc-like patches covered with binding ligands (Figure~\ref{fig:sphere}).
Due to their relatively large micron-size scale, the binding kinetics of these cells are significantly different from the core-shell nano-crystal interactions, which are applicable at much smaller scales~\cite{Duval2008}.
The next few subsections detailed aspect of this model.

\subsection{Binder kinetics} \label{subsec:BK}
Figure~\ref{fig:sphere} illustrates the motion of two deformed spheres, with an identical size of radius~\(R\) (when undeformed).
The spheres are moving in a fluid undergoing planar shear flow.
To simplify the visualisation of the dynamics, consider a moving frame,~$(R_m)$, with origin~$O_m$, fixed on the surface of the sphere~$S_2$ at a point equidistant from the edge of the separation gap.
The unit vectors for this frame of reference are~\(\vec e_z\), \(\vec e_y\) and~\(\vec e_z\).
For a given spatial point ${\vec x} = (x, y, z)$ in this moving frame, the velocity of the fluid is~$G z\vec e_x$, where $G$~is the shear rate.
The total relative velocity (of the sphere~S$_1$ with respect to the sphere~S$_2$) in this frame is ${\vec V}(G, {\vec x}) = U_x {\vec e}_x + U_z {\vec e}_z$\,, where $U_x$ and~$U_z$ are, respectively,   shearing flow (along the plane perpendicular to the line joining the centres), and the velocity of the squeezing motion of the spheres (along the line joining the centres). 
Let~$D(x)$ be the distance between the two spheres.
Define $A_{\text{Tot}} g({\vec x},t)\, d A$ as the number of bonds that are attached between the surfaces~$d A$ at time~$t$ where $A_{\text{Tot}}$~is the total number of binding ligands.
In established research on colloids, the function~$g$ is synonymous with the term \emph{sticking probability}.
The total number of bonds formed is $\int_{A_c} A_{\text{Tot}} g({\vec x},t)\, d A$\,, where~$A_c$ is the area of adhesion (\S\ref{subsec:AA} details  the derivation of this area). 
%
%

The forward and reverse reaction rates for the ligand binding are then written as Boltzmann distributions, allowing highly stretched bonds to be readily broken by thermal energy fluctuations.
The kinetics are also influenced by the surface potential of the two charged surfaces.
Further, we cater for the ligands tilting by a finite angle~$\alpha_0$ with respect to the vertical direction.
This tilt is again expressed as a Boltzmann distribution,~$\mathcal{D}(\alpha_0)$, such that a bond may form between the two spheres for a given angle $\alpha_0 \in (-\frac{\pi}{2}, \frac{\pi}{2})$.
With these degrees of freedom, the bond attachment\slash detachment rates are 
\begin{align}
K_{\text{on}} ({\vec x}) &= K_{\text{on,eq}} \exp \left[ \frac{- \lambda_s(L({\vec x})-{l}_0)^2 + W(D(x))}{2{k}_B T} \right ]\mathcal{D}(\alpha_0), \nonumber \\
K_{\text{off}} ({\vec x}) &= K_{\text{off,eq}} \exp \left[ \frac{ (\lambda_0 - \lambda_s)(L({\vec x})-{l}_0)^2 + W(D(x)) }{2{k}_B T} \right], \label{eqn:reaction_rates}
\end{align}
where ${k}_B$ is the Boltzmann constant, \(T\) is the temperature, $l_0$~is the mean rest length of the binders, $\lambda_0$~is the binder stiffness coefficient, and $\lambda_s$~is the spring constant of the transition state used to distinguish catch ($\lambda_0 < \lambda_s$) from slip ($\lambda_0 > \lambda_s$) bonds~\cite{Dembo1988}. $W(D)$~is the total surface potential described in \S\ref{subsec:LRI}.
In further description of the model we  denote \(D(x) \equiv D\), without loss of generalisation.
As depicted in Figure~\ref{fig:sphere}, $L = \sqrt{D^2 + |{\vec x}|^2}$ is the length of a bond in a stretched configuration.
The energy associated with tilting a bond from its vertical position is~$(1/2)\lambda_{\theta} \alpha_0^2$, ($\kappa_{\theta}$ being the torsion constant) and~\cite{Reboux2008}
\ben
\mathcal{D}(\alpha_0) = \exp\left(-\frac{\lambda_{\theta}\alpha_0^2}{2{k}_B T}\right) \frac{1}{D_0}\,, \quad \alpha_0 = \tan^{-1}\frac{x}{D}\,, \label{eq:tilt}
\een
where $D_0 = \int^{{\pi}/{2}}_{-{\pi}/{2}} \exp\big[{-\frac{\lambda_{\theta}\alpha^2_0}{2{k}_B T}}\big]\, d \alpha_0$ is the normalization constant for all possible tilt orientations along the flow-direction.
In the limit of small binding affinity and abundant ligands on the binding surface (i.e., ${A_{\text{Tot}} K_{\text{on, eq}}}/{ K_{\text{off, eq}} } \ll 1$), the bond ligand density evolves in accordance with the \textsc{pde}~\cite{Dembo1988, Hodges2002, Reboux2008}
\ben
\frac{d g}{d t} = A_{\text{Tot}} K_{\text{on}} - K_{\text{off}} g\,, \quad g = 0 \quad  \text{for }   x \ge R_c \,,
\label{eq:coll_fac}
\een
where the material derivative $\frac{d g}{d t} = \frac{\partial g}{\partial t} + {\vec V} \cdot \nabla g$\,.
%
%
%

%
\subsection{Long range interactions}\label{subsec:LRI}
Derjaguin, Landau, Verwey and Overbeek theory is utilized to describe the interaction between the charged cell surfaces as well as due to the ions dispersed in the fluid medium, via a surface potential~$W(D)$ (equation~\eqref{eqn:reaction_rates}).
Only the effects of Coulombic repulsion and Van der Waals attraction are incorporated.
Other interactions including hydration effects, hydrophobic attraction, short range steric repulsion, and polymer bridging, which are absent in the length scales of our interest, are neglected~\cite{Gregory2006}.
For two charged spheres, with an identical size of radius~$R$, the potential due to the Coulombic forces in the gap of size~$D$ is 
\ben
W_{\text{C}}(D) = 2\pi \epsilon_0 \epsilon \psi^2_0 R e^{-\delta D} ,
\een
where $\delta$ is the Debye length, $\epsilon$ and~$\epsilon_0$ are the dielectric constant of vacuum and the medium, respectively, and $\psi_0$~is the average \emph{zeta potential} or the electric potential of the diffuse cloud of charged counterions.
The potential due to the Van der Waal forces for these spherical cells in the regime of close contact  is 
\ben
W_{\text{VW}}(D) = -\frac{A R}{12 D} \quad \text{for }D \ll R\,,
\een  
where \(A\) is the Hamaker constant measuring the van der Waal `two-body' pair-wise interaction for macroscopic spherical objects.

\subsection{Micro-scale forces: bond mechanics} \label{subsec:MicF}
Consider one individual bond formed between the points~O$_m$ on sphere~S$_2$ (which is also the origin of the frame~$(R_m)$) and \(P\)~on sphere~S$_1$ (Figure~\ref{fig:sphere}).
The instantaneous force it exerts on the two spheres has three components: an extensional force related to bond stretching given by Hooke's law, ${\vec f}_E = \lambda_0(L - {l}_0) {\vec e}_L$\,; a force due to surface-charges, ${\vec f}_C = \nabla W(D) {\vec e}_z$\,; and a torsional force proportional to the angle formed by the bond with the vertical, ${\vec f}_T = (\lambda_{\theta} \frac{\alpha_0}{L}) {\vec e}_{L_{\perp}}$\,, where ${\vec e}_L = -(x {\vec e}_x + D {\vec e}_z)/L$ and ${\vec e}_{L_{\perp}} = (-D {\vec e}_x + x {\vec e}_z)/L$ are the unit vectors tangential and perpendicular to the bond as shown in Figure~\ref{fig:sphere}. The operator, $\nabla$, in the expression of the force due to surface charges, ${\vec f}_C$, denotes the derivative with respect to $D$.
The total micro-scale force, due to each component, arising from all such bonds inside the adhesion area is~\cite{Reboux2008}
\ben
{\vec F}_i({\vec x},t) = A_{\text{Tot}} \int_{A_c} g({\vec x},t) \vec f_i({\vec x},t) \, dA({\vec x}, t), \quad i \in \{ E, C, T\}\\.
\label{eq:Fs} 
\een

\subsection{Macro-scale forces: hydrodynamics} \label{subsec:MacF}
Next, we present the hydrodynamic force resisting the relative motion of two deformable cells moving along their lines of centres (i.e., along the direction~${\vec e}_z$, Figure~\ref{fig:sphere}) as well as the force and couple associated with the transverse translation of these drops along the direction of the flow (i.e., along the direction~${\vec e}_x$), in the Stokes regime.
Haber~\cite{Haber1973} considered a very general problem of two viscous drops with unequal sizes, velocities and viscosities, translating along their lines of centres in bispherical coordinates.
The drops are in close proximity ($D \ll 1$\,, Figure~\ref{fig:sphere}) so that the drag force is derived via lubrication theory.
For the case of two identical drops moving toward each other with equal speed, Haber's solution for the drag force reduces to \cite{Karrila1991}
\ben
{\vec F^*_z} = \frac{{\vec F_z}}{6\pi\mu R U_z} = \frac{2}{3} \sinh \beta \sum^\infty_{n=1} C_n \frac{K^0_n(\beta) + \lambda K^1_n(\beta)}{Q^0_n(\beta)+\lambda Q^1_n(\beta)} {\vec e_z}\,, 
\label{eqn:squeezeForce}
\een
where the functions
\begin{align}
K^0_n(\beta) &= 2\left[(2n+1)\sinh 2\beta + 2 \cosh 2\beta - 2e^{-(2n+1)\beta}\right], \nonumber
\\
K^1_n(\beta) &= (2n+1)^2 \cosh 2\beta - 2(2n+1) \sinh 2\beta 
\nonumber\\&\quad{}
- (2n+3)(2n-1) + 4e^{-(2n+1)\beta}, \nonumber
\\
Q^0_n(\beta) &= 4\sinh(n-\tfrac{1}{2})\beta \sinh(n+\tfrac{3}{2})\beta\,, \nonumber
\\
Q^1_n(\beta) &= 2\sinh(2n+1)\beta - (2n+1)\sinh 2\beta\,, \nonumber
\\
C_n &= \frac{n(n+1)}{(2n-1)(2n+3)}\,.
\end{align}
The starred quantity in equation~\eqref{eqn:squeezeForce} is the non-dimensional counterpart of the force~\(\vec F_z\). 
The product~$(\lambda \mu)$ is the viscosity of the fluid inside the drop ($\mu$~being the viscosity of the fluid outside the drop). 
Parameter~$\beta$ is related to the distance between the centres of the cell by $2 D_c + D = 2R \cosh \beta$ (Figure~\ref{fig:sphere}). 
The scalar $U_z = \|{\vec U_z}\|$ (where $\| \cdot \|$ denotes the magnitude of a vector).
For inviscid cells, we set $\lambda = 0$ in equation~\eqref{eqn:squeezeForce} and, following Cox~\cite{Cox1967}, break the summation into an ``inner sum'' $\sum^N_{n=1}$ and an ``outer sum'' $\sum^\infty_{n=N+1}$ (with the breakpoint~\(N\) determined by requiring $\beta N \sim 1$) and simplify the expression for large~\(N\) to obtain the drag force on either cell,
\ben
{\vec F^*_z} = \frac{{\vec F_z}}{6\pi\mu R U_z} = \left[ \frac{1}{3}\ln \left( \frac{R}{D} \right) + \frac{2}{3}(\gamma + \ln 2) \right] {\vec e_z},
\label{eqn:Fz}
\een
where $\gamma = 0.57722$ is Euler's constant.
The flow in between the inviscid cells, along a plane perpendicular to the line joining the centers of the cell, is not dominated by the gap region and thus lubrication theory does not apply~\cite{Davis1989}.
However, in a slow shear flow regime, we use the leading order results obtained by Zinchenko~\cite{Davis2009} for the forces and the couples:
\ben
{\vec F^*_s} = \frac{{\vec F_s}}{6\pi\mu R U_x} = 1.15 {\vec e_x} \,,\quad {\vec T^*_s} = \frac{{\vec T_s}}{8\pi\mu R^2 U_x} = 1.1 {\vec e_y}\,,
\label{eqn:Fs}
\een
where $U_x = \|{\vec U_x}\|$.
By principle of linear super-position, the total forces and the torques on the two moving cells in slow shear flow conditions are the sum of the contributions from equations~(\ref{eqn:Fz},\ref{eqn:Fs}); ${\vec F} = {\vec F_z} + {\vec F_s}$\,, and ${\vec T} = {\vec T_s}$\,.

\subsection{Non-dimensionalized system} \label{subsec:ND}
We non-dimensionalize the length scales with respect to the undisturbed radius of the cell,~\(R\), the tension on cell surface with a reference tension,~$\tau$, and introduce the following dimensionless variables denoted by stars
\begin{align}
&x = R x^*, \quad D = R D^*, \quad L = R L^* , \nonumber 
\\
&K_{\text{on}} = K^*_{\text{on}} K_{\text{on,eq}}\,, \quad K_{\text{off}} = K^*_{\text{off}} K_{\text{off,eq}}\,, \quad g = g^* K_{\text{eq}} \,, \nonumber
\\
&t = t^*/G \,, \quad U_{x,z} = U^*_{x,z} R K_{\text{off,eq}}\,, \quad \text{and} \quad V^* = U_x^* + U_z^*, 
\end{align}
where $K_{\text{eq}} = A_{\text{Tot}} {K_{\text{on, eq}}}/{K_{\text{off, eq}}}$\,.
Two time-scales are introduced, one associated with the fluid shear rate,~$G^{-1}$, and the other with the rate constant of the binding-unbinding reaction,~$K_{\text{off,eq}}$.
This is done to neglect the lower order terms in the non-dimensional form of equations (and in the limit of slow time-scales), as shown in \S\ref{subsec:MicMac}.
Further, we introduce the following non-dimensional parameters,
\ben
{r} = \frac{\lambda_0 {l}^2_0}{{k}_B T}\,, \quad \lambda^*_{\theta} = \frac{\lambda_{\theta}}{{k}_B T}\,, \quad \lambda^*_s = \frac{\lambda_s}{\lambda_0}\,, \quad \epsilon = \frac{{l}_0}{R} \,.
\label{eqn:param}
\een
The non-dimensional form of the reaction rates, equation~\eqref{eqn:reaction_rates}, bond-density evolution, equation~\eqref{eq:coll_fac}, and the boundary conditions are
\begin{align}
&K^*_{\text{on}} = \exp \Big[ -\lambda^*_s \frac{{r}}{2\epsilon^2} (L^* - \epsilon)^2 - \frac{\lambda^*_{\theta}}{2} \alpha^2_0 + W^*(D) \Big] / P_0 \,, \nonumber
\\
&K^*_{\text{off}} = \exp \Big[ (1-\lambda^*_s)\frac{{r}}{2\epsilon^2}(L^* - \epsilon)^2 + W^*(D) \Big] , \label{eqn:Konoff}
\\
&\frac{G}{K_{\text{off,eq}}} \frac{\partial g^*}{ \partial t^*} + V^* \frac{\partial g^*}{\partial x^*} = K^*_{\text{on}} - K^*_{\text{off}} g^* , \nonumber
\\
&g^* = 0 \quad  |x^*| \ge R^*_c\,, \label{eqn:g*}
\end{align}
where $W^*(D)={W(D)}/({2{k_B}T})$.
Similarly, the non-dimensional form of the micro-scale forces arising from all bonds are 
\begin{eqnarray} 
{\vec F}^*_E(U, D) &=& - \int_{A^*_c} g^* (1 - 1/L^*) \big[ x^*{\vec e}_x + D^*{\vec e}_z\big] \, d A^*, \nonumber
\\
{\vec F}^*_C(U, D) &=& -(1 / {l}_0 {r}) {\vec e}_z \int_{A^*_c} g^* \left[\pi \epsilon_0 \epsilon \psi^2_0 \kappa \frac{2R_1 R_2}{R_1 + R_2}\left(e^{-\kappa D}-\frac{A}{24 D^2} \right)\right]  \, d A^*, \nonumber
\\
{\vec F}^*_T(U, D) &=& (\kappa^*_{\theta} / {r}) \int_{A^*_c} (g^*/L^{*2}) \alpha_0 \big[ -D^*{\vec e}_x + x^*{\vec e}_z\big] \, d A^*, \label{eq:microscale_nd}
\end{eqnarray}
where ${\vec F}^*_E$, ${\vec F}^*_C$ and ${\vec F}^*_T$ are the non-dimensional forces due to \emph{extension}, \emph{surface charges} and \emph{torsion}, respectively, and ${\vec F}^*_i = {\vec F}_i / (A_{\text{Tot}}K_{\text eq} \kappa_0 R^3)$.


\subsection{Adhesion area: $A_c$} \label{subsec:AA}
The adhesion of the two cells occur inside a circular patch of area of $A_c=\pi R^2_c$\,, where $R_c$~is the radius of the patch (as shown in Figure~\ref{fig:sphere}).
To determine this radius, we split the computational domain into two distinct regions: an inner region (which is the gap between the two cells) and an outer region, outside the gap (in the horizontal sense).
We define the variables~$(p, c)$ as the excess pressure inside the cell and the surface tension in the inner region, respectively.
The variables~$(P, C)$ are the corresponding variables in the outer region. 
The membrane curvature is denoted by~\(Q\).
Inside the inner region, the cells (with undisturbed circular radius~\(R\), Figure~\ref{fig:sphere}); deforms under the action of adhesive forces and the shear flow.
For slow shear rates ($G\le5$\,s$^{-1}$), Jensen found that the stress balance on one of the cell surface (Figure~\ref{fig:sphere}), at the leading order in the separation gap between the two cell surfaces, is~\cite{Hodges2002}
\ben
A_{\text{Tot}}K_{\text{on}}({\vec x}) \left( {\vec f}_E + {\vec f}_C + {\vec f}_T \right) + p_0 {\vec e}_z = Q c^0 {\vec e}_z - c_x^0 {\vec e}_x\,,
\label{eqn:LOInner}
\een
whereas the stress balance, at the next order of approximation, is
\ben
p_1 {\vec e}_z = Q c^1 {\vec e}_z - c_x^1 {\vec e}_x + \mu G {\vec e}_x\,, 
\label{eqn:FOInner}
\een
where the superscripts, \(0\) and~\(1\), denote variables at the leading order and the next order approximation, respectively, and the variable~$c^1_x$ denote derivative of the surface tension in the moving frame.
The cell membrane is flat (with~\(Q\)=0) along the majority of the inner region, except at the edge of the gap where it connects with the outer region (point K, Figure~\ref{fig:sphere}).
The leading order and the next order approximation in the stress balance on the cell surface, in the outer region, is
\begin{align}
P^0 {\vec e}_R &= Q C^0 {\vec e}_R - C^0_{\theta} {\vec e}_{\theta}\,, \label{eqn:LOOuter}
\\
P^1 {\vec e}_R &= Q C^1 {\vec e}_R - C^1_{\theta} {\vec e}_{\theta} + \mu G {\vec e}_x\,, \label{eqn:FOOuter}
\end{align}
where ${\vec e}_R$ and~${\vec e}_{\theta}$ are the unit vectors along the directions normal and tangential to the cell surface, and the subscript~$\theta$ denotes the derivative of a quantity along a tangent to the surface.

For an undisturbed cell (satisfying stress balance~\eqref{eqn:LOInner} at leading order), Jenson~\cite{Hodges2002} found an asymptotic expression for the patch radius in the limit of small separation gap, with no sharp corners and with a uniform tension and curvature in the outer region of the cell.
In the slow shear rate regime, we account for the first order correction to the surface tension and the net pressure (equations~(\ref{eqn:FOInner},~\ref{eqn:FOOuter})), match the solution in the inner and the outer region with matching conditions at the juncture (point~K, Figure~\ref{fig:sphere}), and derive the expression of the patch radius
\ben 
R_c = R \left | \sqrt{1 - \left[1- \frac{\epsilon \mathcal{M}}{r}\left( 1+ \frac{\lambda^*_\theta}{r} \right) \right]^2} - \frac{\mu G}{\tau} - W^*(D) \right |, 
\label{eqn:Xc}
\een
where $\mathcal{M}={\lambda_0 A_{\text{Tot}} K_{\text{on,eq}}R^2\epsilon}/{\tau}$ ($\tau$ is the uniform tension of the undisturbed cell) is a dimensionless parameter related to the strength of the cell surface; that is, reducing~$\mathcal{M}$ corresponds to making the cell less deformable and vice-versa.
Finally,  for neutral cells in static equilibrium ($G=\lambda^*_{\theta}=W^*(D)=0$), the expression of the patch radius derived by Jenson~\cite[relation 2.10]{Hodges2002} is recovered from~\eqref{eqn:Xc}. 



\subsection{Micro-Macro coupling} \label{subsec:MicMac}
Under the assumption that the bonds are formed and broken at a rate sufficiently rapid for them to remain in equilibrium, i.e., ${G}/{K_{\text{off,eq}}} \ll 1$\,, the unsteady binding effects (i.e., the time-dependent term in equation~\eqref{eqn:g*}) are neglected and the evolution equation for the sticking probability is solved at steady-state
\ben 
V^* \frac{\partial g^*}{\partial x^*} = K^*_{\text{on}} - K^*_{\text{off}} g^*, \label{eq:g_quasi}
\een
whose steady-state, analytical solution is 
\ben
g^*(x^*) = \frac{1}{V^*} \int^{R_c}_{x^*} K^*_{\text{on}}(s_2) \exp\left[{-\frac{1}{V^*}\int^{s_2}_{x^*}K^*_{\text{off}}(s_1)\,d s_1}\right] d s_2\,. \label{eq:Maineqn}
\een
The coupling between the macro-scale and the micro-scale description is obtained via the global force balance on the spheres in the horizontal and vertical directions, and the torque balance about the center of mass of the spheres.
Assembling the forces and couples from the fluid hydrodynamics, equations~(\ref{eqn:Fz}, \ref{eqn:Fs}), and the total forces due to bond-extension, surface charges and bond-torsion, arising from all bonds, equation~\eqref{eq:Fs}, we obtain
\begin{eqnarray}
&&(\vec F^*_E + \vec F^*_C + \vec F^*_T) \cdot {\vec e}_z +  F_z^* = 0\,, \nonumber
\\
&&(\vec F^*_E + \vec F^*_C + \vec F^*_T) \cdot {\vec e}_x + F_s^* = 0\,, \nonumber
\\
&&\frac{R}{2} (\vec F^*_E + \vec F^*_C + \vec F^*_T) \cdot {\vec e}_x + T_s^* = 0\,, \label{eqn:balance}
\end{eqnarray} 
where the factor~${R}/{2}$ denotes the effective radius for a system of two spheres of equal radius. $F_z^*, F_s^*, T_s^*$ are the scalar values of the forces and couples outlined in equations~(\ref{eqn:Fz}, \ref{eqn:Fs}). The micro-macro force balance is utilized to solve for the unknowns, the translational speeds $U_x$ and~$U_y$ and the separation gap between the spheres~$D$.
Finally, equations~(\ref{eqn:Konoff}, \ref{eq:microscale_nd}, \ref{eq:Maineqn}, \ref{eqn:balance}) form the system of equations that fully describes the binding kinetics of two deforming spheres moving in a slow shear flow fluid conditions.
The next section describes numerical results and the biophysical implications of this system.

\section{Binder kinetics at steady state} \label{sec:results}
Table~\ref{tab:parameter} lists the parameters used in our numerical calculations.
The parameter values are chosen so that they closely replicate the adhesion-fragmentation of neutrophiles in slow viscous shear flow conditions.
For example, the \textsc{p}-selectine molecule extends about \(40\)\,nm from the endothelial cell membrane, so when combined with its ligand \textsc{psgl}-1 it is reasonable to take ${l}_0 \approx 100$\,nm as an estimate of the length of the unstressed bond~\cite{Shao1998}.
Typically, neutrophils have a size of $R \approx 4\,\mu$m which gives the length ratio $\epsilon \approx 0.025$ (equation~\eqref{eqn:param}).
Hichmuth~\cite{Shao1998} measured variations of up to three orders of magnitude \emph{in vivo} in measuring the values of the  microvillus stiffness,~$\lambda_0$, as well as the membrane tension of an undisturbed cell.
Direct measurements of the parameters,~$A_{\text{Tot}}$, $K_{\text{on,eq}}$ and~$K_{\text{off,eq}}$ are scarce, although values in several thousands have been used in previous models~\cite{Hammer1996}.
Since we do not wish to study the effects of finite rotation of the ligands~\cite{Reboux2008} or the effect of catch-versus-slip bonds~\cite{Dembo1988}, the corresponding parameters related to these material properties are fixed at $\lambda^*_{\theta}=1.0$ and $\lambda^*_s=0.5$\,, respectively.
The dielectric constant in vacuum is $\varepsilon_0 = 8.854 \times 10^{-12}$\,Farad\,m$^{-1}$, whereas the permittivity of water at temperature~$25^\circ$C is $\varepsilon=78.5$ (not to be confused with~$\epsilon$ which is a length ratio, equation~\eqref{eqn:param}).
The dissolved salt (furnishing the ions in the fluid) is assumed to be a 1-1 electrolyte with a zeta potential of $\psi_0=25$\,mV (corresponding to the surface potential studies by Gregory~\cite[Chap.~3]{Gregory2006}).
We assume that the solute concentration in the fluid only effects the Debye length,~$\delta$.
The Boltzmann factor is taken as ${k}_BT=4 \times 10^{-21}$\,J. 
\begin{table}
\centering
\caption{Parameters common to all numerical results and used in  studies of the system of equations~(\ref{eqn:Konoff}, \ref{eq:microscale_nd}, \ref{eq:Maineqn}, \ref{eqn:balance}).}\label{tab:parameter}
\begin{equation*}
\begin{array}{ccccc}
\hline
\text{Parameter} & \text{Value} & \text{Units} & \text{Source}\\
\hline
A_{\text{Tot}} & 10^9 & \text{m}^{-2} &\cite{Hammer1996} \\
K_{\text{on, eq}} & 10^2 & \text{s}^{-1} &\cite{Hammer1996} \\
K_{\text{off, eq}} & 10^{14} & \text{s}^{-1} &\cite{Hammer1996} \\
\lambda_0 & 10^{-5} \text{ -- } 10^{-2} & \text{N\,m}^{-1} &\cite{Mani2012} \\
\mu & 10^{-3} & \text{N\,s\,m}^{-2} &\cite{Reboux2008} \\
G & 1\text{ -- }5 & \text{s}^{-1} &\cite{Reboux2008} \\
l_0 & 10^{-7} & \text{m} &\cite{Shao1998} \\
R & 4 \times 10^{-6} & \text{m} &\cite{Shao1998} \\
\tau & 2.5 \times 10^{-5} \text{ -- } 2.5 \times 10^{-3} & \text{N\,m}^{-1} & \cite{Shao1998} \\
\hline
\end{array}
\end{equation*}
\end{table}
%
%

We numerically solve the multi-scale model using the adaptive Lobatto quadrature (via Matlab function \verb|quadl|) to evaluate the integral in equation~\eqref{eq:Maineqn}, which is then coupled with the system of algebraic equations, equation~\eqref{eqn:balance}, to calculate the unknown macroscale speeds at steady state, $U_x(G)$ and~$U_z(G)$, the separation distance,~$D$, and, subsequently, the sticking probability,~$g^*(U_x, U_z, D)$. 

\begin{figure}
\centering
\includegraphics[scale=0.6]{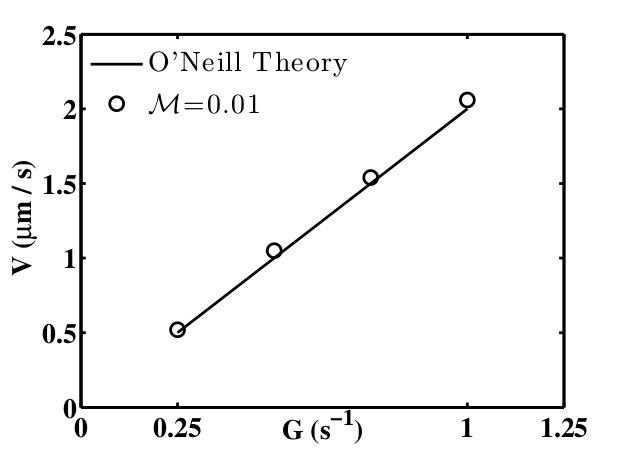}
\caption{Hydrodynamic speed of two identical, hard spheres of same size and at the same separation distance along the line joining the centres~($\circ$) with the results calculated using the theory of O'Neill and Majumdar~\cite{ONeill1970}.}\label{fig:Fig2}
\end{figure}

As a preliminary step, the model was validated by estimating the net hydrodynamic speed, $V=\|{\vec U}_x+ {\vec U}_z\|$, of two noninteracting, nearly rigid spheres as a function of shear rate.
For this purpose, spherical capsules with a membrane stiffness coefficient of nearly rigid spheres, $\mathcal{M}=0.01$\,, and high ligand stiffness, $\lambda_0=10^{-2}$\,N\,m$^{-1}$, was used.
Simulations indicate that the hydrodynamic velocity increases linearly with shear rate from~$0.5\,\mu$m/s at~$0.25\,\text{s}^{-1}$ to~$2\,\mu$m/s at \(1.0\)\,s$^{-1}$ (Figure~\ref{fig:Fig2}).
These values are in excellent agreement (with $< 0.1 \%$ difference in $l^{\infty}$-norm) with the velocity calculated by O'Neill and Majumdar~\cite{ONeill1970} for the motion of two hard spheres of the same size and at the same separation distance in a linear shear field.

\begin{figure}
\centering
\subfigure[$g^*(x^*=0.5)$ versus \(G\) at $\lambda_0=10^{-3}$\,N\,m$^{-1}$, $\delta=1.0$]{\includegraphics[width=0.48\linewidth]{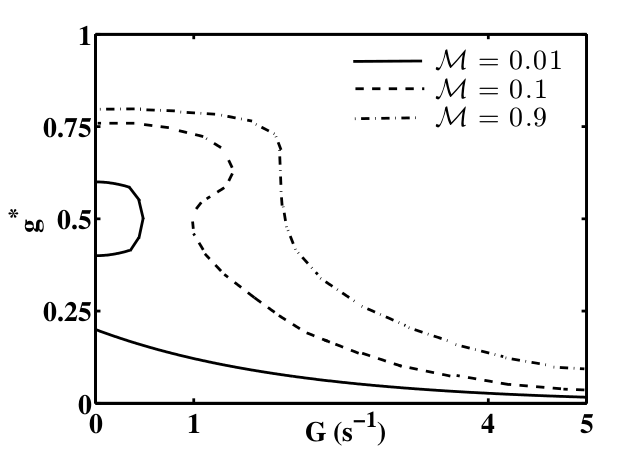}}
\quad
\subfigure[$g^*(x^*=0.5)$ versus \(G\) at $\mathcal{M}=0.01$\,, $\lambda_0=10^{-3}$]{\includegraphics[width=0.48\linewidth]{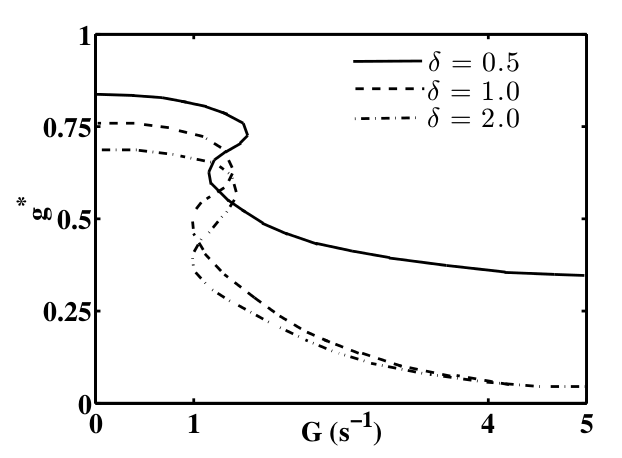}}
\quad
\subfigure[$g^*(x^*=0.5)$ versus \(G\) at $\mathcal{M}=0.01$\,, $\delta=1.0$]{\includegraphics[width=0.48\linewidth]{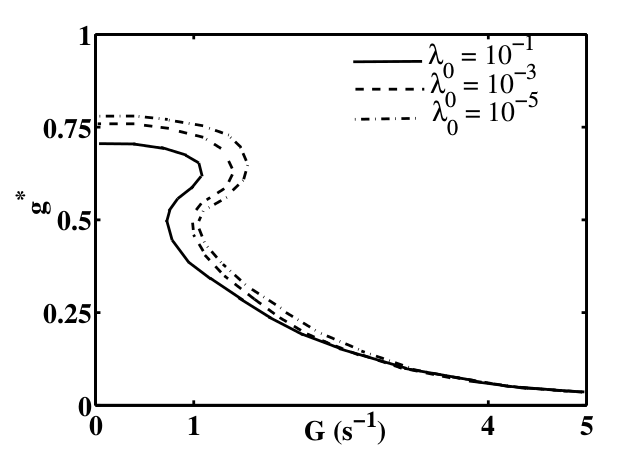}}
\caption{Steady state transition curves of the sticking probability, $g^*(x^*=0.5)$, versus shear rate,~\(G\), for different (a)~membrane stiffness,~$\mathcal{M}$, (b)~screening lengths,~$\delta$, and (c)~binder stiffness,~$\lambda_0$.
Three different adhesion-fragmentation transitions are detected when the membrane stiffness is changed: (1)~continuous reversible transition (dash-dot curve); (2)~continuous reversible transition (dashed curve); and (3)~discontinuous irreversible transition (solid curve).
No qualitative changes in the transition curves are observed if the screening length or the binder stiffness is changed, within the limits listed in Table~\ref{tab:parameter}.}\label{fig:Fig3}
\end{figure}

\begin{figure}
\centering
\subfigure[Hydrodynamic speed at $\delta=1.0$]{\includegraphics[width=0.48\linewidth]{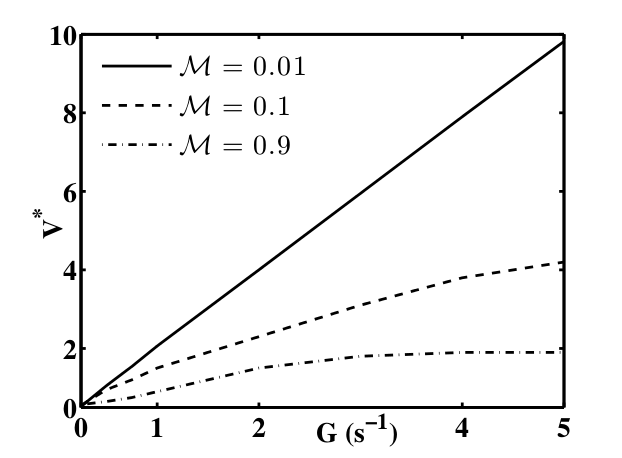}}
\quad
\subfigure[Hydrodynamic speed at $\mathcal{M}=0.1$]{\includegraphics[width=0.48\linewidth]{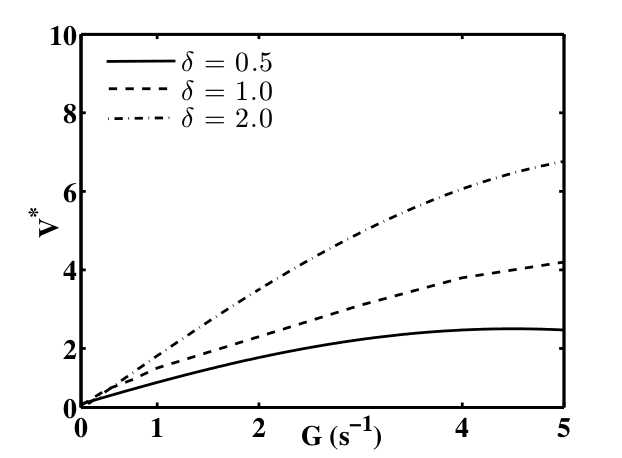}}
\caption{Effect of (a)~cell surface deformability and (b)~fluid ionic conditions, on the hydrodynamic speed of the cells in shear flow.
The material parameter for these simulations is fixed at $\lambda_0=10^{-3}$\,N\,m$^{-1}$.}\label{fig:Fig5}
\end{figure}

Next, we explored the flow/binding kinematics of the deforming spheres in a uniform shear flow.
Figure~\ref{fig:Fig3}a depicts the steady-state solution in the \emph{sticking probability}-shear flow ($g^*,G$) phase space at a horizontal distance $x^*=0.5$ from the origin of the moving frame, and for variable surface deformabilities,~$\mathcal{M}$.
An \emph{adhesion} phase is defined when the majority of the binders (inside the adhesion area,~$A_c$) are hooked with each other, i.e., $g^* > 0.5$\,; otherwise the spheres are in the \emph{fragmentation} phase.
A third, \emph{bistable} phase, in which the spheres exhibit a stable steady-state adhesion and fragmentation, simultaneously coexists on the phase plane.
Figure~\ref{fig:Fig5} presents the boundaries of the adhesion and fragmentation region which are computationally tracked as a continuation of the limit points of~$g^*$.
In another numerical experiment, we found that changing the limits of the values of~$g^*$ which defines these regions, has very little impact on the boundaries of these regions.
The results of these experiments are not shown here for conciseness. 

In the present study, three different types of adhesion-fragmentation kinematics are found.
For example, for nearly rigid cells (Figure~\ref{fig:Fig3}a, $\mathcal{M}=0.01$), the transition from adhesion to fragmentation phase (and vice-versa) is irreversible and discontinuous.
For this curve, the adhesive effects are strong for low shear rates (i.e., $g^*$ has a stable steady-state branch with $g^* > 0.5$ in the shear rate range $G < 0.5$s$^{-1}$).
As the shear rate increases to the critical value, $G=0.5$\,, the system abruptly jumps to a steady-state value in the fragmentation phase (i.e., $g^* < 0.5$) and remains in this phase even if the fluid shear rate is reduced below this critical value.
For deformable cells (Figure~\ref{fig:Fig3}a, $\mathcal{M}=0.1, 0.9$), this transition is reversible with flow, and either changes continuously ($\mathcal{M}=0.9$ curve) or discontinuously through the bistable region ($\mathcal{M}=0.1$ curve).

Figure~\ref{fig:Fig3}b,~c, respectively, presents the effects of the different ionic conditions in the surrounding fluid affecting the screening length,~$\delta$, and the binder stiffness coefficient, on the flow-kinematic phase space.
Strong surface adhesion is observed in highly ionic fluids (i.e., fluids represented by shorter screening lengths,~$\delta$, Figure~\ref{fig:Fig3}b) and with elastic binders (i.e., binders with lower stiffness coefficient, Figure~\ref{fig:Fig3}c).
A shorter screening length implies a smaller separation distance between the interacting surfaces, and hence a strong adhesion.
Similarly, elastic binders aid bond formation which favors surface adhesion.
Another observation is the absence of any qualitative differences within the curves in Figure~\ref{fig:Fig3}b,c, a finding which is consistent with previous theoretical predictions~\cite{Hammer1996}. 
%
%

\begin{figure}
\centering
\subfigure[Magnitude of the total microscale forces]{\includegraphics[width=0.48\linewidth]{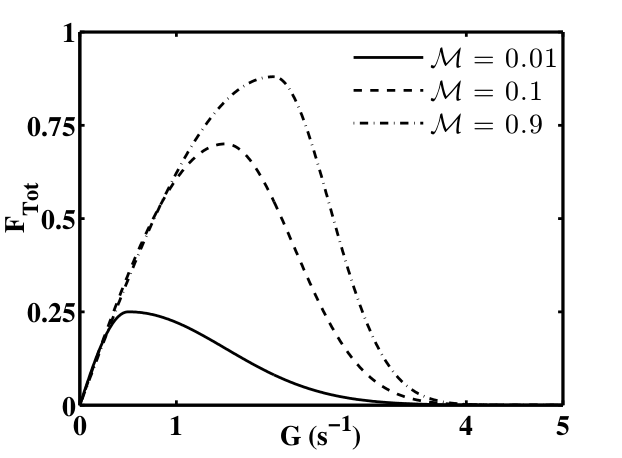}}
\quad
\subfigure[Minimum vertical separation distance]{\includegraphics[width=0.48\linewidth]{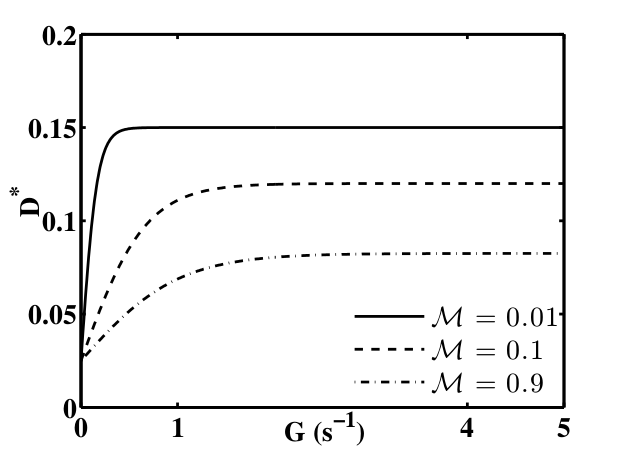}}
\caption{Total microscale force,~$F_{\text{Tot}}$, and the separation distance,~$D^*$, at a horizontal distance, $x^*=0.5$ from the origin of the reference frame, versus the fluid shear rate,~$G$.
Material parameters for these simulations are fixed at $\lambda_0=10^{-3}$\,N\,m$^{-1}$ and $\delta=1.0$\,.
The nonlinear relation between the total microscale forces and the fluid shear rate is due to a nonlinear dependence of the forces on the minimum separation distance, which is a variable.}\label{fig:Fig4}
\end{figure}

Physically, the abrupt hysteretic transitions in the \emph{sticking probability} between the adhesion and the fragmentation regimes (i.e., the transition curves in Figure~\ref{fig:Fig3}) is explained by the relation between the magnitude of the total micro-scale binding force, $F_{\text{Tot}}=\|{\vec F}^*_E+{\vec F}^*_C+{\vec F}^*_T\|$, and the fluid shear rate (Figure~\ref{fig:Fig4}a).
In general, at low non-dimensional shear rates ($G < 1.0$) the total force due to the stretching and tilting of the ligands along the flow direction increases with the shear rate.
However, in strong flow conditions ($G > 2.0$), the bonds rupture and there is a rapid decay in the total binding force.
As the shear rate is increased, the bound ligands are unable to prevent some degree of fragmentation between the two surfaces and start to yield.
Consequently, the total adhesive forces decrease, eventually leading to a state where the cells are free from nearly all adhesive bonds.
However, the strength of the total microscale forces, depends on the membrane surface tension (i.e., the stiffness coefficient,~$\mathcal{M}$).

The nonlinear relation between the microscale forces and the fluid shear rate as well as the cell deformability is tentatively justified as follows.
With increasing shear rate the binders are advected away from the vertical alignment, the \(z\)-component of the torsion force,~$\vec F_T$, as well as the surface force,~$\vec F_C$, pushes the cells farther away.
However, for sufficiently large separation distances, the bonds stretch and the extension forces,~$\vec F_E$ ($\propto D^*$), tend to pull the cells close to each other.
All these forces depend on the minimum separation, $D^*=D^*(G, \mathcal{M}, \delta)$, (Figure~\ref{fig:Fig4}b), which varies nonlinearly with fluid shear rate, cell surface deformability and the ionic conditions in the fluid, and thus account for the non-linear variation versus the separation distance.

Further, we investigated the effects of the cell surface deformability (Figure~\ref{fig:Fig5}a) as well as the Debye length (Figure~\ref{fig:Fig5}b) on the hydrodynamic speed of the cells, $V^*=\|{\vec U}^*_x + {\vec U}^*_z\|$.
At nearly zero shear rate, the hydrodynamic speed does not vary significantly with the deformability coefficient,~$\mathcal{M}$.
In contrast, pronounced differences were observed at higher shear rates.
In particular, the hydrodynamic speed for nearly rigid cells ($\mathcal{M}=0.01$\,, Figure~\ref{fig:Fig5}a) and cells immersed in weakly ionic fluids ($\delta=2.0$\,, Figure~\ref{fig:Fig5}b) increased appreciably.
Conversely, only a modest increase in the hydrodynamic speed of more complaint cells ($\mathcal{M}=0.01, 0.1$ curves in Figure~\ref{fig:Fig5}a) or cells immersed in strong electrolytic solvent ($\delta=0.5, 1.0$ curves in Figure~\ref{fig:Fig5}b) occurred with increasing~shear.

Altogether, cell deformation induced by the hydrodynamic forces due to fluid flow modulates the ligand-mediated cell adhesion kinetics.
Deformable cells (i.e., cells with higher stiffness coefficient,~$\mathcal{M}$) exhibit compact binding with a higher magnitude of the total microscale binding forces, (Figure~\ref{fig:Fig4}a), remain closer to each other (Figure~\ref{fig:Fig4}b) and move slowly (Figure~\ref{fig:Fig5}a). 
Dri and colleagues \cite{NDri2003}  attributed these features due to an increased adhesion contact area (equation~\eqref{eqn:Xc}) as well as the reduction in the overall magnitude of hydrodynamic forces in the gap between the cells (hydrodynamic forces are proportional to the hydrodynamic speed in slow viscous fluid-flow limit, Figure~\ref{fig:Fig5}a) experienced by the more complaint cells. 
Since the total hydrodynamic forces are proportional to the speed, a reduced hydrodynamic speed in a highly ionic aqueous environment (comparing the curves in Figure~\ref{fig:Fig5}b) results in a strong surface adhesion, an observation corroborated with experimental findings \cite{Gregory2006}.


%
\begin{figure}
\centering
\subfigure[$\lambda_0$-$G$ phase plane at $\delta=1.0$]{\includegraphics[width=0.48\linewidth]{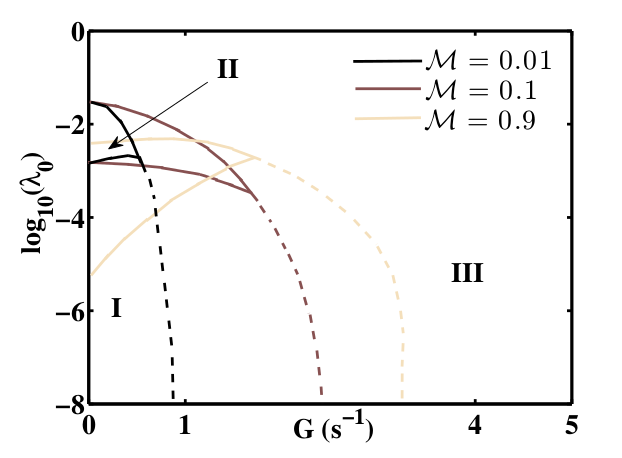}}
\quad
\subfigure[$\lambda_0$-$G$ phase plane at $\mathcal{M}=0.1$]{\includegraphics[width=0.48\linewidth]{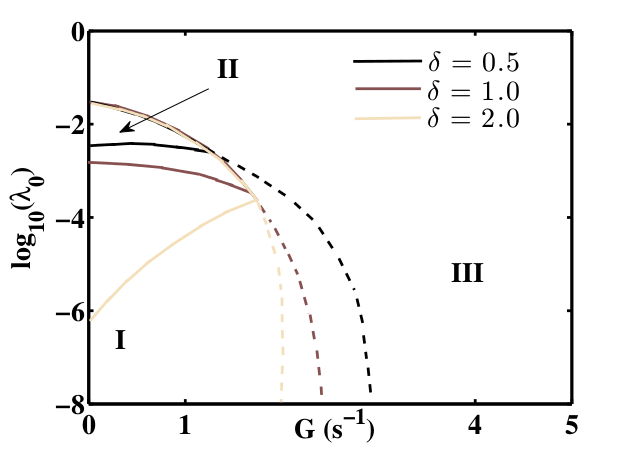}}
\caption{$\lambda_0$-$G$ phase plane highlighting regions of adhesion~(I), bistabity~(II) and fragmentation~(III) for different (a)~cell surface stiffness coefficient,~$\mathcal{M}$, and (b)~Debye length,~$\delta$, at a horizontal distance, $x^*=0.5$\,, from the origin of the reference frame.
The adhesion\slash fragmentation regions are characterized by $g^* > 0.5$ and $g^* < 0.5$\,, respectively.
The boundary of the bistable region is the locus of the limit points of the hysteretic $g^*$-$G$ curves in Figure~\ref{fig:Fig3}.}\label{fig:Fig6}
\end{figure}

Figure~\ref{fig:Fig6} identifies the domain of adhesion (region~I), bistability (region~II) and fragmentation (region~III), within a select range of materials parameters used in our numerical calculations (Table~\ref{tab:parameter}).
Bistability is an intrinsic property of any biophysical system exhibiting hysteretic transitions, such as the adhesion-fragmentation transitions shown in Figure~\ref{fig:Fig3}.
As the flow shear rate increases from zero, the initially attached cell surfaces, detach at a critical shear rate (i.e., $g^*$ drops below 0.5).
If the shear rate decreases below this critical value the process is reversed, then cells surfaces reattach (i.e., the value of~$g^*$ rises above~\(0.5\)) but at a critical shear rate different than the previous threshold.
Figure~\ref{fig:Fig6} highlights, with solid lines, the locus of all such~$(g^*,G)$ critical-points enclose the bistable region in the material parameter space.
The dashed lines correspond to the nullcline $g^*=0.5$\,.

Cell adhesion bistability occurs from a tug-of-war between two kinetic processes taking place within the contact area, bond formation which aids adhesion and bond rupture~\cite{Lauffenburger1993}. 
As seen in Figure~\ref{fig:Fig6}, the factors affecting adhesion are low fluid shear rate and elastic binders (i.e., lower stiffness coefficient,~$\lambda_0$) which assists bond formation, deformable membrane surface (or larger value of the membrane stiffness coefficient,~$\mathcal{M}$) which leads to increased attachment area and lower the magnitude of total hydrodynamic force (proportional to the hydrodynamic speed, Figure~\ref{fig:Fig4}a), and strong ionic conditions (i.e., lower screening length,~$\delta$) which reduces the separation distance between the cell surfaces.

Bistability has been reported in a variety of experiments, especially those involving cell-wall and cell-cell adhesion.
Brunk and Hammer~\cite{Brunk1997} detected bistability in an \emph{in vitro} set-up of cell-free assay characterized by a single bond type (\textsc{e}-selectin and its ligands), mimicking rolling neutrophils over stimulated endothelial surface.
Yago et al.~\cite{Yago2002} gave further evidence of bistability via numerical simulations of neutrophils rolling on a carbohydrate selectin-ligand substrate under flow---a phenomenon later corroborated by King~\cite{King2005}.
%
%


\section{Conclusions and discussion} \label{sec:end}
Section~\S \ref{sec:model} presented a new unified, exhaustive, multi-scale model for the adhesion of two spherical, deforming cells via tiltable, elastic ligands in an ionic fluid subject to a homogeneous shear flow.
Section~\S \ref{sec:results} demonstrated that the transition between the adhesion and the fragmentation phases can be reversibly continuous, reversibly discontinuous, or irreversible, depending on the deformability of the cell surface, the strength of the ionic fluid medium and the stiffness of the binding ligands.
In particular, deformable cells exhibit strong adhesion.
We attributed this partly due to the increased cell-cell contact area as well as reduction in the magnitude of the hydrodynamic forces experienced inside the gap between the cells.
Strong ionic fluid conditions favor adhesion through lowering of the hydrodynamic forces as well as reduction in the cell separation gap.
A bistable region signifying the coexistence of both aggregation and fragmentation domains, was numerically detected for a select range of material and fluid parameters (Figure~\ref{fig:Fig6}).

Although the proposed model is able to describe key features in cell adhesion, several issues still need to be addressed \cite{Reboux2008}.
For example, nonlinearity of the micro-scale forces can significantly modify the micro-macro hydrodynamic force balance thereby modifying the adhesion region.
Our approach also excludes spatial inhomogeneity arising through the material parameters, the effects of catch behavior ($\kappa^*_s > 1.0$), non-equilibrium binding effects, stochasticity and the discrete number of bonds~\cite{Zhu2000}, the cellular viscoelasticity (needed to fully describe the cell rheology~\cite{Dembo1988}), the electro-viscous drag on the spherical surfaces surrounded by ionic solution~\cite{Jia2006} (which modifies the fluid velocity across the channel between the cells), as well as shearing forces large enough to tear the binding ligands from their anchoring surface~\cite{Varenberg2007}.
All these effects can lead to several non-trivial behavior (including the possible absence of hysteretic behaviour in flow-phase transition) that deserves a full numerical investigation in the near future.

\paragraph{Acknowledgement} This work was supported in part by the Adelaide University startup funds and the Australian Research Council Discovery grant DP150102385.
We thank Dr.~Edward Green and Dr.~Trent Mattner in the Department of Mathematical Sciences, Adelaide University, for providing useful insights at various stages of model development.

\bibliography{mainbib}

\end{document}